# Functionalized CdTe Fluorescence Nanosensor for the Sensitive Detection of Water Borne Environmentally Hazardous Metal Ions

Pooja and Papia Chowdhury [*]

Department of Physics and Materials Science and Engineering, Jaypee Institute of Information Technology, Noida 201309, Uttar Pradesh, India.

*Corresponding author:papia.chowdhury@jiit.ac.in

**Highlights**

- The selectivity and optical sensing ability of COOH-CdTe fluorescence nanosensor for the detection of different water borne environmentally hazardous metal ions have been tested.
- Variation of temperature and time influences the optical response of CdTe fluorescence nanosensor.
- For the first time, the efficient optical sensing efficiency for CdTe nanosensor is tested by its fluorescence "Turn-Off" response for water dissolved hazardous metal ions in between concentration range from $10^{-14}$M to $10^{-6}$M.
- A very efficient sensing from real samples like: river water, paint water and rain water near industrial areas have been successfully verified by CdTe nanosensor by Principal component analysis and fluorescence testing.

**Abstract**

In this work, we have investigated the possibility of using COOH functionalized CdTe quantum dots (QDs) as fluorescent nanosensor for the detection of some environmentally hazardous metal ions ($Cr^{3+}$, $Pb^{2+}$, $Cu^{2+}$, $Zn^{2+}$ and $Co^{2+}$) in aqueous phase. COOH-CdTe QDs are prepared in aqueous medium and characterized by UV-visible, photo luminescence and FTIR spectrometry. Optical response of CdTe nanosensor is observed to be time and temperature dependent. A strong "Turn-Off" fluorescence response of CdTe is observed in presence of all metals ions ($X^{n+}$). The output of Stern volmer relation, graphical Job's plot and FTIR data established the existence of strong 1:1 complexation ($X^{n+}$: $^-$OOC-CdTe) betweenCdTe and metal ions. CdTe shows its efficient sensing ability for $Pb^{2+}$, $Cr^{3+}$ and $Cu^{2+}$ ions up to concentration $10^{-14}$ M which was established by its low detection of limit (LOD)validation. Further, strong appearance of $F-F_0/F_0$ establishes the better sensing capability of CdTe QDs for $Cr^{3+}$, $Pb^{2+}$, $Cu^{2+}$ ions which are far better than the existing fluorescent sensors available in industry. The proposed optical techniques and principal component analysis of sensing were successfully applied for testing of dissolved metal ions in real samples like: paint, river water, rain water etc which validates the applicability of COOH functionalized CdTe nanosensor as a effective successful optical sensor for water dissolved metal ions which are responsible for water pollution.

1. Introduction

From past few decades, nanometric materials have been attracting a great deal of interest among the researchers due to their variable size, structure and electronic properties which significantly different from their bulk counterparts [1]. Among different nanomaterials, semiconductor quantum dots (QDs) have drawn much attention towards researchers due to their tremendous applicability in different areas of optical, electronic and medical industries [2]. The novel and unique properties of QDs arise from their variable sizes and composition [3]. According to the law of quantum confinement, as the particle size of QDs decrease, there is discreteness in the electron and hole energy states [4], and therefore the band gap comes out to be a function of the QD's diameter. QDs are mainly composed of semiconductor materials form II-VI, III-V and IV-VI groups of the periodic table. PbS, CdTe, CdSe, GaAs, lnP etc are some well known industrically available QDs who have several application areas like solar cell [5], LED[6], quantum computing [7], medical imaging [8,9] and fluorescent marker [10,11] etc. QDs are also turn out to be efficient optical sensors for the detection of many diseases [12], environmentally polluting reagents like hazardous gases[13], polluting materials in air [14] and water [15].

Like many other reasons of environmental pollution areas like: air, food, sound etc, water pollution is one of the biggest concerned area of civic society which we are dealing right now. Release of undesirable and pollutant materials like organic and inorganic, medical and radioactive wastages are the major reason for water pollution. Hazardous metal ions from chemical, leather and medical industries are one of the main source of water pollution which mix up into river, lake etc water through drainage and leads to the contamination of water. The polluted water further enters the biological system through drinking or bathing and causes different diseases [16]. The industrial wastages, mainly include water dissolvable metal ions like $Cr^{3+}$, $Pb^{2+}$, $Cu^{2+}$, $Zn^{2+}$, and $Co^{2+}$[17]. $Cr^{3+}$, $Pb^{2+}$ and $Cu^{2+}$ are the most toxic metallic pollutants which can cause many disease like memory loss, anemia, lung cancer, low white blood cell count and neurological problems[18]. As minerals some metal ions like $Cr^{3+}$, $Pb^{2+}$, $Cu^{2+}$, $Zn^{2+}$, and $Co^{2+}$ are needed for healthy life and should be found in drinking water. $Cr^{3+}$ and $Cu^{2+}$ are the main required minerals to our body. Deficiency of $Cr^{3+}$ and $Cu^{2+}$ can cause some diseases like muscle weakness, anaemia, inhibition of protein synthesis and heart disease. According to WHO guideline the minimum concentration of required metals as minerals in ground water are: 0.0005 mg/L ($9.8 \times 10^{-9}$M) for $Cr^{3+}$ [19] , 0.005mg/L ($2.8 \times 10^{-8}$M) for $Pb^{2+}$ [20] and 1mg/L ($1.5 \times 10^{-5}$M) for $Cu^{2+}$ [21]. But if the concentration increases beyond a certain limit, the excessive existence of these ions may cause many health hazards and can increase the pollution level of water. WHO's recommendation about the permissible limit for dissolved metal ions in ground water are $1.8\times10^{-7}$M for $Cr^{3+}$, $4.8\times10^{-8}$M for $Pb^{2+}$ and $1.9\times10^{-5}$M for $Cu^{2+}$[22]. According to WHO report, in India (Delhi/NCR region) due to the presence of many industries around Ganga/Yamuna rivers, the concentration of $Cr^{3+}$, $Pb^{2+}$ and $Cu^{2+}$ are found to be $7.8\times10^{-6}$M, $4.8\times10^{-7}$M and $2.2\times10^{-5}$M [22]. Similarly in urban region around Beijing city the concentration of these ions are found to be $9.8\times10^{-7}$ M, $9.8\times10^{-8}$M and $3\times10^{-5}$ M [23], which are definitely harmful for living body and environment. Directly or indirectly large amount of human/animal population used to consume the river water for drinking and bathing purpose.Therefore, fast and efficient detection of hazardous metal ions in aquatic environment is very much needed.

Many scientific techniques are used in today's industry for detection of metal ions like chromatography [24], atomic fluorescence spectrometry [25], atomic spectroscopy [26] etc. Similarly many sensing systems are available like electrochemical sensors[27], Bio sensor[28], organic fluorophores [29] etc. Most of the above mentioned methods are time consuming, costly [30]. Similarly most of the mentioned sensors require highly precise sample, turn around time and expensive equipment to operate [31]. In industry, organic fluorophores like: phycoerythrin, allophycocyaninare the mostly used sensors for metal ions as they detect the sensing systems by very fast optical signals [32]. Organic fluorophore sensors are largely in use in industry due to their low detection limit [33] and low cost, but problem with most of them is their less quantum yield (QY)(25-50%)[34]. So there is a requirement of some low cost, optical sensors with high QY and low detection limit. QDs can be the answer for that. There are many QDs like CdTe, CdSe, GaAs, lnP etc are available in industry which are used as optical metal sensors due to their high QY [34] and low detection limit [35]. CdTe is the mostly applicable crystalline nanoparticle and used as optical sensor due to its extreme ability of tuning the optical signal in the visible spectral range (380-740 nm) by controlling its diameter [36,37]. Over the last few years CdTe QDs have been widely in use as fluorescent biological markers [36,38] and as metal ion detector [39,17]. They are also been used in medical imaging probe [40] and LED construction [41]. Due to their photochemical stability and high QY (35-70%), CdTe QDs have been chosen to be a good alternative to most available organic fluorophores [42]. But due to their inorganic nature like all QDs, CdTe nanoparticle are also insoluble in water and toxic in nature [43]. So handling of CdTe QDs as optical sensor is not very effective in their original form.Capping reagents can trap the QDs by eliminating their surface defects and improving the photochemical stability as well as solubility of QDs in water[17]. Capping reagent can also remove the impurities from the surface of QD and prevent agglomeration [44,45]. Normally,organic capped QDs are widely used for sensing different water dissolved metal ions and also detoxify heavy metal ions by binding with them through polymerization [45,46]. There are many reported works on CdSe, CdS, CdSeS/ZnS QDs with different capping reagents like 2-merceptoethanol ($C_2H_6OS$) [47], thioglycolic acid ($C_2H_4O_2S$) [48], glutathione ($C_{10}H_{17}N_3O_6S$) [17]. But capping reagents have some drawbacks. The photoluminescence (PL) efficiency of QDs were observed to decreased [17] due to long chain structure of the capping reagents. CdTe with capping agents like 2-merceptoethanol [49], citric acid ($C_6H_8O_7$) [50], glutathione [51] etc show decrease in PL efficiency. So to increase the PL efficiency of CdTe, small sized capping reagent like carboxyl ($\geq$C=0), amino ($\geq$N-H) or aldehyde ($\geq$C=O-H), functional group [52] etc can be tried. Presence of carboxyl (-COOH) group can make CdTe ionized, less toxic [53] and water soluble. Thioglycolic acid is used to produce COOH functionalized CdTe QD [54].The carboxylic end promotes a negative charge at the outer layer that avoids QDs aggregation and in turn increases PL efficiency. Presence of carboxylic groups also assure the stability of QDs due to charge repulsion [54]. They also participate in reactions with various environments. Till now no work has been reported about the sensing ability of COOH funtionalized CdTe QDs.

In the present work, we have reported the selectivity and optical sensing ability of COOH-CdTeQDs for the detection of many water borne environmentally hazardous metal ions such as $Cr^{3+}$, $Pb^{2+}$, $Cu^{2+}$, $Zn^{2+}$ and $Co^{2+}$. COOH-CdTe QDs show their low detection limit for $Pb^{2+}$, $Cr^{3+}$ and $Cu^{2+}$ detection with a very low concentration $10^{-14}$ M which is far better than the existing sensors which can be used to detect water dissolved hazardous metal ions. Our existing research will definitely benefit the society by designinga effective optical metal ion QD nanosensor.

## 2. Experimental

### 2.1 Materials

COOH functionalized CdTe QDs were purchased from Sigma Aldrich in powder form. Deionized water (millipore) was used for solution preparation. Cupric chloride ($CuCl_2$), zinc chloride ($ZnCl_2$), lead acetate trihydrate ($Pb(CH_3CO).3H_2O$), cobalt nitrate hexahydrate ($Co(NO_3)_2.6H_2O$), chromium nitrate ($Cr(NO_3)_3$) were purchased from Sigma Aldrich. Firstly, a primary solution of COOH-CdTe QDs stirred properly until the solution became transparent by maintaining 0.7mg/ml ratio. Finally the secondary solution have been prepared to maintain the concentration of COOH-CdTe QDs as $3\times10^{-5}$ M.

### 2.2 Apparatus

The UV–Visible absorption spectra were recorded at 300K and by varying temperature from30º C to 70ºC by a Perkin Elmer spectrophotometer (model Lambda-35) with a varying slit width in the range 190–900 nm. All luminescence measurements were made with a Perkin Elmer spectrophotometer (Model Fluorescence-55) with a varying slit width (excitation slit = 10.0 nm and emission slit = 5 nm) ranging from 260 to 900 nm. The Model LS 55 Series uses a pulsed Xenon lamp as a source of excitation. Deionized water (Millipore) was used for measuring absorption and emission spectra of capped QDs. Fourier transform infrared (FTIR) spectra were recorded with Perkin-Elmer FTIR spectrophotometer (model spectrum BX-II source: nichrome glower wire with DTGS detector) ranging from 400 to 4000nm. All optical measurements were performed under ambient conditions. Principle component analysis (PCA) was completed by Matlab.

## 3. Results and discussion

### *3.1 UV-Vis absorption and emission analysis*

CdTe QDs is insoluble in hydroxylic medium. COOH capping provides good water solubility to CdTe QDs in aqueous medium. Presence of –COOH functional group can also help CdTe to bio-conjugate with different organic or inorganic sensing systems which in turn may exhibit a good optical sensing capability of COOH-CdTe QDs. Functionalized CdTe QDs shows UV-Vis absorption spectrum covering the range from 250 nm to 600 nm with a sharp peak at 288 nm and a weak shoulder around 488 nm[55]. The peak at 288 nm is due to the presence of carboxyl functional group($>C=0$)[56] in aqueous medium (Figure 1a). The absorption band at 488 nm is arised from $1S(h) \rightarrow 1S(e)$ in CdTe QDs, which is due to the transition of an electron from the valence band (VB) to the conduction band (CB). This excitonic band (VB-CB) of COOH-CdTe in aqueous medium at 488 nm is related to the size of QD [57]. The average particle size (2R) of the CdTe QDs can be determined by using the absorption edge of the spectrum using the formula [58]

$$2R_{(CdTe)} = \frac{0.1}{(0.138 - 0.0002345\lambda_C)} \text{nm} \quad (1)$$

Where, 2R is the diameter of the particle and $\lambda_C$ is the absorption edge. The average particle size of CdTe QDs have been estimated as ~ 8.3 nm which is similar as that of the computed particle size of stable $(CdTe)_6$ nanoparticle reported in our previous work [55]. The fluorescence spectrum of COOH-CdTe QDs shows a narrow Stokes shifted emission band between ~500-600 nm [59] with high quantum yield in aqueous medium. The shape of the emission spectrum of CdTe QD is symmetric in nature and with a FWHM of 42 nm which indicates the uniform and monodisperse particle size of dissolved QDs (Figure 1b) [44].

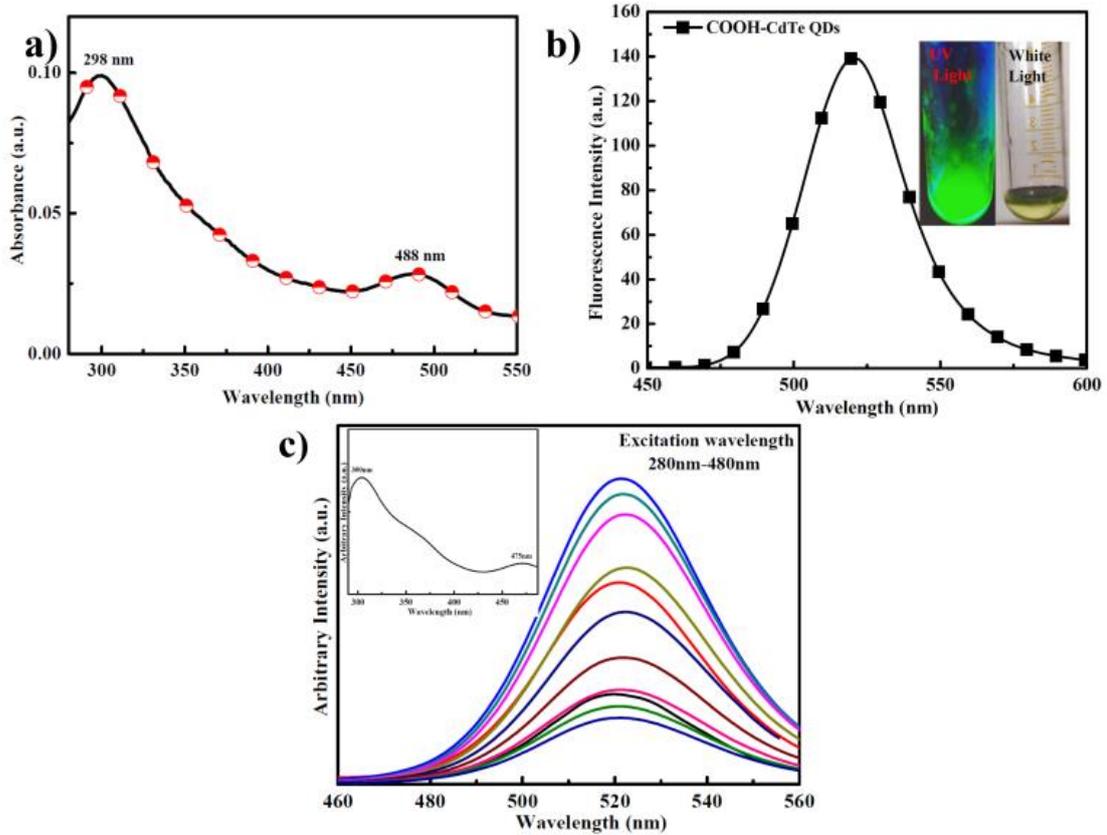

**Figure 1:** (a) UV–Vis absorption spectrum of COOH-CdTe QDs in hydroxylic medium (b) Emission spectrum of COOH-CdTe QDs in hydroxylic medium (c) Emission spectrum of COOH-CdTe QDs with varying excitation wavelength (280-480 nm) in hydroxylic medium inset (excitation spectrum of COOH-CdTe QDs emission wavelength 530 nm).

The average particle size (2R) of COOH-CdTe QD from its luminescence data has been calculated using the following formula [60],

$$D = (9.8127 \times 10^{-7})\lambda^3 - (1.7147 \times 10^{-3})\lambda^2 + (1.0064)\lambda - 194.84 \qquad (2)$$

Where, D=2R is the diameter of the QDs in nm and λ is the fluorescent emission wavelength of CdTe QDs. The particle size of the prepared CdTe QD has been estimated as ~ 8 nm from the luminescence data which perfectly matches with particle size of QD obtained from absorption data and computed data for (CdTe)$_6$ structure [55]. The COOH-CdTeQD solution is pale yellow under white light [39] and emits intense yellow green luminescence under UV light (300nm) which can be seen by naked eyes also (inset of Figure 1b). The above observation is also consistent with the excitation dependent emission spectra of CdTe QD solution obtained by varying excitation wavelengths (280-480 nm) (Figure 1c). The emission characteristic is further verified by the excitation spectrum (inset of Figure 1c).

To identify exact ambient condition and better sensitivity of CdTe QDs as good optical sensor, some main factors governing the optimum synthesis conditions for CdTe QDs are required. The factors may include variable temperature, time, pH etc. Also presence of COOH capping helps CdTe to bio-conjugate with different metal ions which in turn may exhibit sensing capability towards different hazardous metal ions. In the following sections we will discuss the ambient condition for COOH-CdTe QDs to be used as sensor and its sensing capabilities towards different hazardous water borne metal ions in detail.

*3.2 Effect of time and temperature on fluorescence intensity*

Studies on the effect of time and temperatures on fluorescence intensity of functionalized CdTe QDs are very necessary to obtain an optimum condition to develop a sensitive fluorescence sensor. The fluorescence intensity of CdTe nanosensor is checked with variable time period. We observed an enhancement in the fluorescence intensity of COOH-CdTe QDs over a period of a week (Figure 2a). The reason of increase in the fluorescence intensity may be due to the photo-activation of CdTe QDs. However, there is a large quenching in the intensity is observed in fluorescence signal when the QDs were stored for a period of one month in dark room (Figure 2a). No shifting of fluorescence peak is observed throughout the time period. Due to its high fluorescence intensity, the time of one weak is chosen as the best reaction condition for further experiment with CdTe.

Temperature of a solution can also cause a difference in photoluminescence intensity of the nano-sized materials [51]. We have studied the effects on fluorescence intensity of CdTe QD against the variation of temperature in order to determine the optimum working condition for CdTe QDs. Figure 2b shows the emission spectra of CdTe QDs at different temperatures i.e 30°C, 40°C, 50°C, 60°C and 70°C. At 30°C, the fluorescence emission peak (Figure 2b) exhibited enhancement with no shift in peak position. At 50°C, the process of agglomeration of QDs has been started and hence we got decrement in fluorescence intensity with a small red shift in peak position (Figure 2b). At temperature 70°C the QDs completely agglomerated and turned to bulk phase, which is reflected with the diminishing of QDs fluorescence peak (Figure 2b). So 30°C was chosen as the optimal temperature for further experiments.

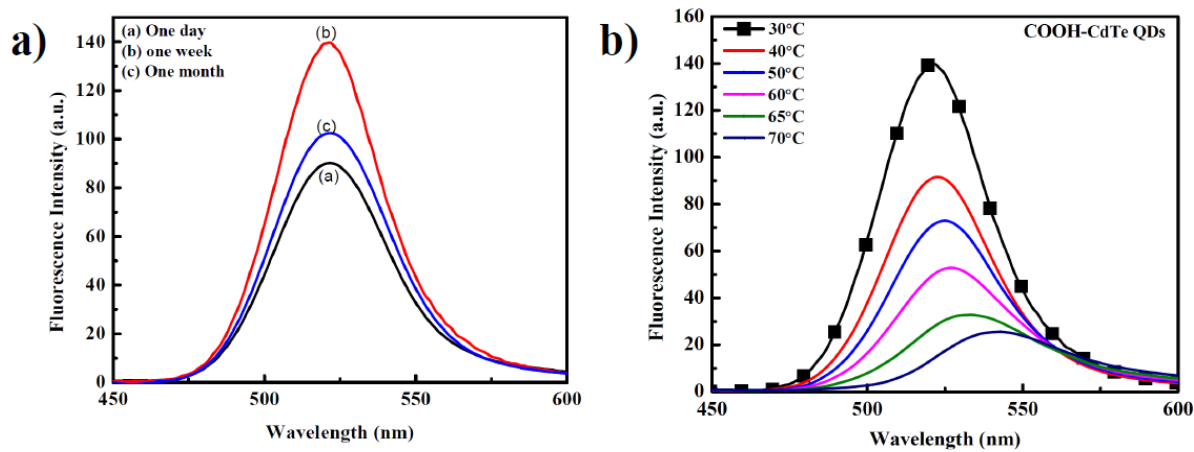

**Figure 2:** Effect (a) time and (b) temperature on the fluorescence intensity of COOH-CdTeQDs in hydroxylic medium.

*3.3 Optical sensing by COOH-CdTeQDs for dissolved hazardous metal ions in water medium*

The UV–vis absorption spectra of COOH-CdTe QDs in the presence of different ions ($Cr^{3+}$, $Cu^{2+}$, $Pb^{2+}$, $Zn^{2+}$ and $Co^{2+}$) were reported in Figure 3a-b and SD1a-c. With the increasing concentrations of ions ($10^{-14}$ M to $10^{-6}$ M), CdTe shows an increase in absorbance at 488nm with a small red shift upto ~498 nm. The red shift in peak position indicates the increase in band gap and hence the increased particle size of QDs in presence of metal ions [61]. The increase in particle size may be due to the complex formation between functionalized CdTe QDs and metal ions by the replacement of internal hydroxyl ion with external metal ion. The above observation has been further verified by the photoluminescence data of CdTe in presence of different metal ion concentration.

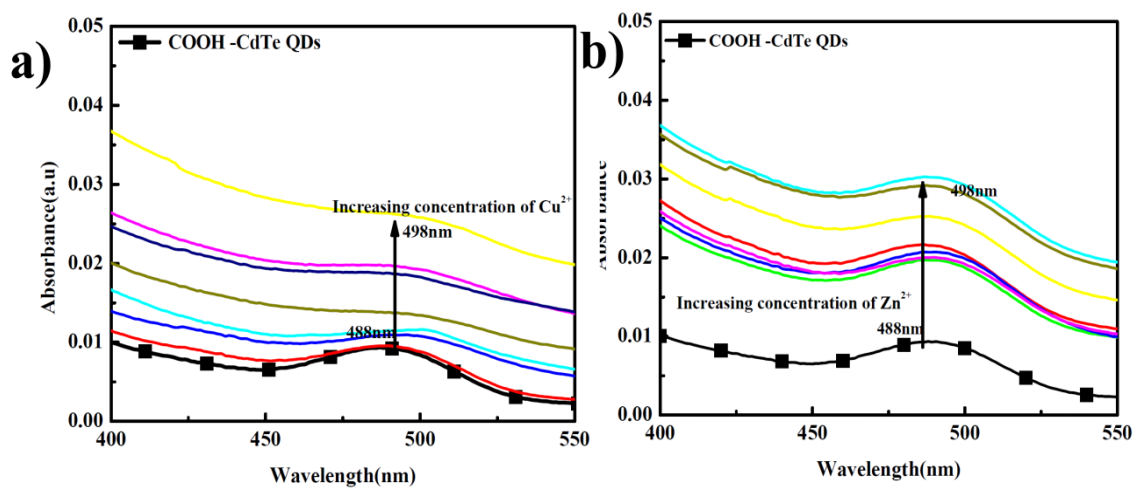

**Figure 3:** UV–Vis absorption spectra of COOH-CdTe QDs in presence of increasing concentration for **a)** $Cu^{2+}$ ions (1.6 x $10^{-14}$ M to 0.12 x $10^{-6}$ M) **b)** and for $Zn^{2+}$ ions (0.12 x $10^{-6}$ M to 1 x $10^{-6}$ M) in hydroxylic medium.

Fluorescence response of COOH-CdTe QDs in the presence of different water dissolved hazardous heavy metal ions ($Cr^{3+}$, $Pb^{2+}$, $Cu^{2+}$, $Zn^{2+}$ and $Co^{2+}$) were observed to be different (Figure 4a,b and SD 2a-c). The concentrations of $Cr^{3+}$, $Pb^{2+}$, $Cu^{2+}$, $Zn^{2+}$ and $Co^{2+}$ metal ions have been used in the range between $1.6 \times 10^{-14}$ M to $0.12 \times 10^{-6}$ M but $Zn^{2+}$ and $Co^{2+}$ did not show any response in the concentration range $1.6 \times 10^{-14}$ M to $0.18 \times 10^{-7}$ M. We have chosen the concentration range for $Zn^{2+}$ and $Co^{2+}$ ions from $0.12 \times 10^{-6}$ M to $1 \times 10^{-6}$ M (Figure 4b, SD 2c). A stable fluorescence "turn off" response with small red shift were observed in the emission profile of COOH-CdTe QDs in presence of different metal ions.

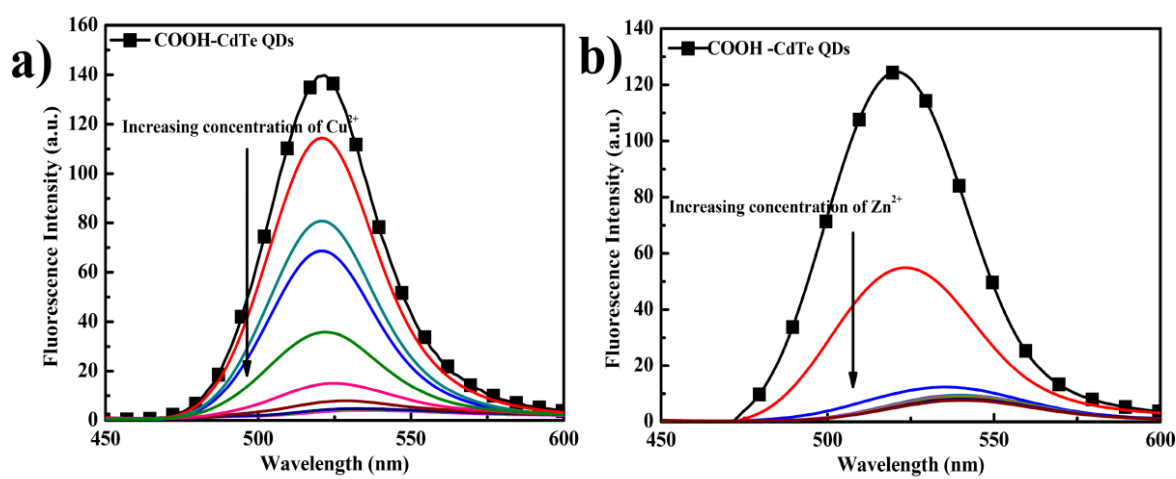

**Figure 4:** Emission spectra of COOH-CdTe QDs in water medium and in the presence of different increasing concentration for **(a)** $Cu^{2+}$ ($1.6 \times 10^{-14}$ M to $0.12 \times 10^{-6}$ M) and for **(b)** $Zn^{2+}$ ($0.12 \times 10^{-6}$ M to $1 \times 10^{-6}$ M) metal ions.

The red shifted decrement in the fluorescence intensities in presence of different metal ions validated the complex formation ($X^{n+}$: $^-$OOC-CdTe) between metal ions ($X^{n+}$) and carboxylic group present on the surface of CdTe QDs. The complex formation would be possible due to the replacement of H from the carboxyl group. The incorporation of $X^{n+}$ ions into the QDs results in the surface defects of QDs leading to a nonradiative recombination of the excitons and fluorescence quenching. With the increasing concentration of metal cations, the reduction of $X^{n+}$ to $X^+$ may appear by conduction band electrons as $X^{n+}$ ions which can accept the electrons from conduction band to become $X^+$. With the presence of more $X^+$ ions, the QD surface become more crowded, then some $X^+$ ions would replace the H atom from the QD surface to make the complex ($X^{n+}$:$^-$OOC-CdTe) formation. As a resut of complex formation, the luminescence of QDs are observed to be quenched. The newly formed $X^{n+}$: $^-$OOC-CdTe complex structure has a lower energy level than the energy level of HOOC-CdTe QD.

So a red shift in fluorescence emission were also observed for ⁻OOC-CdTe QDs in presence of metal ions. The whole mechanism is described in scheme 1a-c.

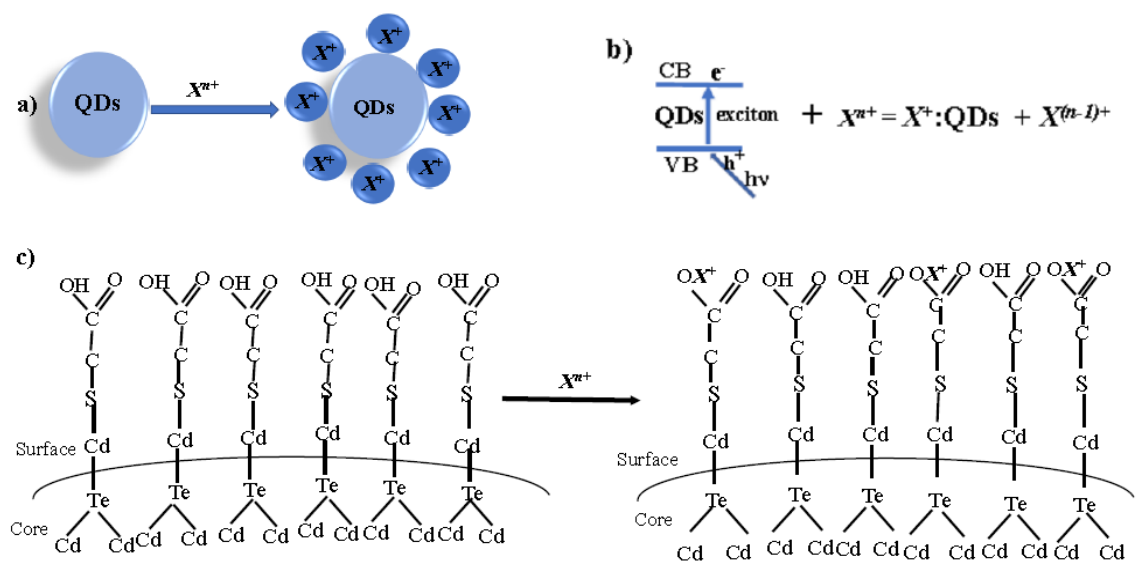

**Scheme 1: a)** Schematic illustration for quenching process of ⁻OOC-CdTe QDs with $X^{n+}$ ions. **b)** Schematic of fluorescent sensing of $X^{n+}$ as electron acceptor from conduction band (CB). **c)** Schematic illustration of surface structure of ⁻OOC-CdTe QDs and replacement of H by $X^{n+}$ metal ions.

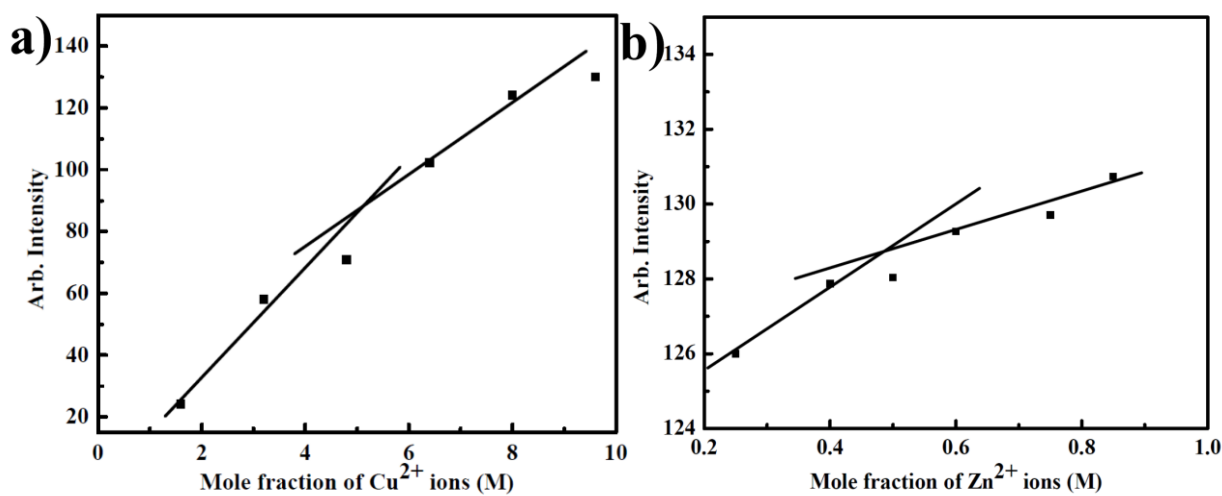

**Figure 5:** Job plot between COOH-CdTe QDs and (a) $Cu^{2+}$, (b) $Zn^{2+}$ metal ions.

Further to verify the stiochiometry of complex formation we have used the job plot method between QDs and metal ions [62]. Job plot data represents the variation of fluorescence emission of COOH-CdTe with the mole fraction of the metal ions as sensing probe. The intersection of the linear portions of the Job plot approximately provided the binding stoichiometry of sensing probe and sensor [63]. The observed value of mole fraction of $Cr^{3+}$, $Pb^{2+}$, $Cu^{2+}$ at 5 M (Figure 5a, SD 3a,b) and $Zn^{2+}$, $Co^{2+}$ metal ions at 0.5 M (Figure 5b, SD 3c) validate the 1:1 complexation between these metal ions and COOH-CdTe QDs (Figure 5a,b, SD 3a-c).

*3.4 Sensitivity measurement of functionalized CdTe nanosensor*

To check the sensitivity of proposed COOH-CdTe QDs as sensor for metal ions ($X^{n+}$), the Stern Volmer methods were used. In this method, the variation fluorescence intensities of proposed sensor with the variable concentrations of different metal ions ($X^{n+}$) can be described by Stern-Volmer equation [64] as follows:

$$\frac{I_0}{I} = 1 + K_{sv}[Q] \qquad (3)$$

Where, $I_0$ and $I$ are the emission intensities of COOH-CdTe QDs in the absence and in presence of quencher Q (concentrations of different metal ion ($X^{n+}$)) and $K_{sv}$ as the Stern Volmer constant which also act as the sensitivity constant. Lower the concentration of metal ion means higher the value of $K_{sv}$, and so lower is the detection limit of the sensor for a specific sensing material. High $K_{sv}$ means the sensor is highly sensitive for the particular metal ion[35]. Stern Volmer plots of COOH-CdTe QDs as sensor for different metal ions as sensing medium having various concentrations were observed to be nonlinear in nature and presented in figure 6 a,b and SD 4a-c. The slope of the Stern Volmer plot is obtained from the linearly fitted range of $\frac{I_0}{I}$ versus [Q], which provides the value of $K_{sv}$ (inset of Figure 6a and SD 4a,b). The linearly fitted range was obtained for the concentration range of $1.6 \times 10^{-14}$ M to $4.8 \times 10^{-11}$ M for ions $Cr^{3+}$, $Pb^{2+}$, $Cu^{2+}$ (Figure 6a, SD 4a,b) and of $0.12 \times 10^{-6}$ M to $1 \times 10^{-6}$ M for $Zn^{2+}$ and $Co^{2+}$ (Figure 6b, SD 4c). For $Cu^{2+}$ ions, the Stern-Volmer plot (inset of Figure 6a) shows an excellent linearity with a good correlation coefficient ($R^2 = 0.99$) which gives a maximum value of sensitivity constant ($K_{sv}$) as $30.86 \times 10^{12}$ $M^{-1}$(Table1). This value is far better than the existing $Cu^{2+}$ ion CdTe sensor for the concentration ($0.04 \times 10^{-9}$ M) like TGA-CdTe, LGSH-CdTe [65].

Similarly for $Cr^{3+}$ ions a good linearity is obtained with $R^2$ value as 0.98 (inset of Figure 6b), with the sensitivity constant ($K_{sv}$) as $12.2 \times 10^{12}$ $M^{-1}$ (Table 1). Similar results was observed for $Pb^{2+}$ good sensitivity constant ($K_{sv}$) as $81.9 \times 10^{11}$ $M^{-1}$ (Table 1). Above results show that the COOH-CdTe QDs can act as a highly sensitive sensor for $Cu^{2+}$, $Pb^{2+}$ and $Cr^{3+}$ metal ions compared to the CdTe sensors.

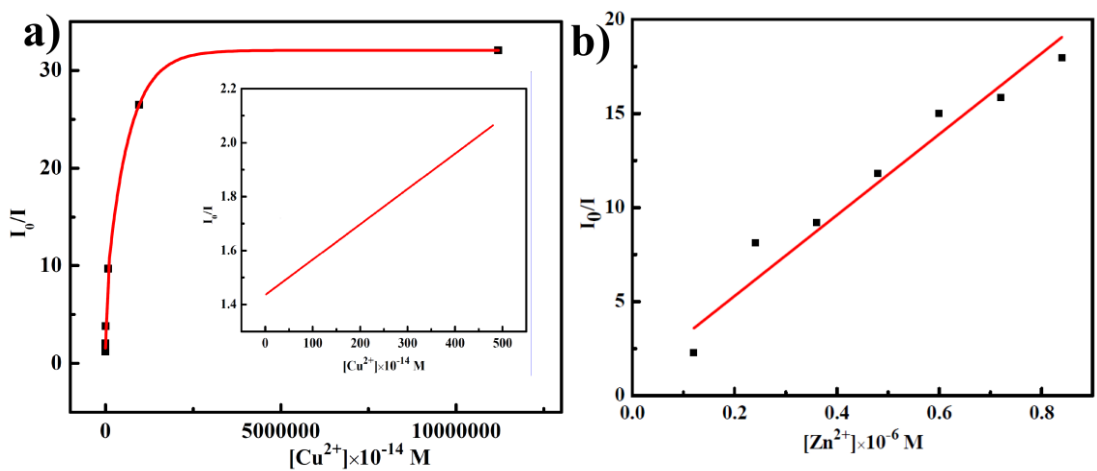

**Figure 6:** Stern Volmer curve for COOH-CdTe QDs with **a)** $Cu^{2+}$, **b)** $Zn^{2+}$ metal ions. Inset the figure linear relationship of $Cu^{2+}$ metal ions.

Further to check the applicability of COOH-CdTe QDs as an active optical metal ion sensor for metal ions, the concept of relative fluorescence has been used. The ratio of change in fluorescence intensity and initial intensity of probe sensor is known as relative fluorescence intensity ($F$-$F_0$/$F_0$), where $F_0$ and $F$ are the initial fluorescence of sensor and fluorescence of complex respectively. The relative fluorescence intensity of $X^{n+}$:⁻OOC-CdTe complex in the presence of different concentrations of metal ions were observed in terms of negative response values: 27, 11, 30, 14 and 17 for $Cr^{3+}$, $Pb^{2+}$, $Cu^{2+}$, $Zn^{2+}$ and $Co^{2+}$ metal ions respectively (Figure 7). Appearance of higher value of relative fluorescence intensity validate COOH-CdTe as better sensitive sensor for $Cr^{3+}$ and $Cu^{2+}$ ions.

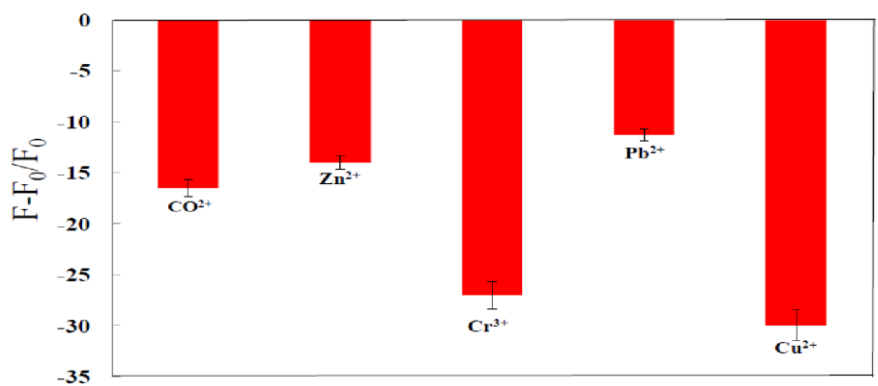

**Figure 7:** Relative fluorescence intensity (F-$F_0$/$F_0$) of COOH-CdTe QDs for metal ions: $Cr^{3+}$, $Pb^{2+}$, $Cu^{2+}$, $Zn^{2+}$ and $Co^{2+}$.

## 3.4 Detection limit for proposed sensor

The lowest concentration of a dissolved metal ion that can be detected at a specified level of confidence (99%) is known as limit of detection (LOD). LOD have been calculated using the equation given below[66]:

$$LOD = \frac{3\sigma}{K} \qquad (4)$$

Where K is the stern volmer constant and σ is the standard deviation of the blank measurements.

| S. No. | Metal ion | Sensitivity ($K_{sv}$) in $M^{-1}$ | Detection limit in M |
|---|---|---|---|
| 1 | $Cr^{3+}$ | $12.2 \times 10^{12}$ | $18.9 \times 10^{-14}$ |
| 2 | $Pb^{2+}$ | $81.9 \times 10^{11}$ | $30.6 \times 10^{-14}$ |
| 3 | $Cu^{2+}$ | $30.86 \times 10^{12}$ | $4.46 \times 10^{-14}$ |
| 4 | $Zn^{2+}$ | $21.94 \times 10^{6}$ | $1.048 \times 10^{-6}$ |
| 5 | $Co^{2+}$ | $12.52 \times 10^{6}$ | $3.2 \times 10^{-6}$ |

**Table 1**: Sensitivity constant ($K_{sv}$) and Detection limit for different ions ($Cr^{3+}$, $Pb^{2+}$, $Cu^{2+}$, $Zn^{2+}$, and $Co^{2+}$).

Under the optimal condition, the calibration curves for the different metal ions ($Cr^{3+}$, $Pb^{2+}$, $Cu^{2+}$, $Zn^{2+}$, and $Co^{3+}$) were presented in Figure 6a-b and SD 4a-c. The LOD of different heavy metal ions are obtained as $18.9 \times 10^{-14}$M for $Cr^{3+}$, $30.6 \times 10^{-14}$M for $Pb^{2+}$, $4.46 \times 10^{-14}$M for $Cu^{2+}$, $1.048 \times 10^{-6}$ M for $Zn^{2+}$ and $3.2 \times 10^{-6}$M for $Co^{2+}$.

## 3.5 The binding mechanism of COOH-CdTe QDs with different metal ions

To check the interaction mechanism of different metal ions with COOH-CdTe nanosensor, Fourier Transformed Infrared (FTIR) spectroscopic technique can be used. FTIR spectra of free COOH-CdTe QDs and $X^{n+}$: $^-$OOC-CdTe complexes have been analyzed in hydroxylic environment (Figure 8a-f). Free COOH-CdTe QDs show a wide and very strong IR absorption band between 1600-2100 cm$^{-1}$ region due to the symmetric stretching mode of carboxylic ($\nu_{COOH}$) group and strong and wide band between 2100-3000 cm$^{-1}$ due to hydroxyl ($\nu_{OH}$) stretching mode (Figure 8a). A weak peak at 832 cm$^{-1}$ is observed for the C-S ($\nu_{CS}$) stretching mode (not shown in Figure 8), which remains unaltered in presence of different metal ions in COOH-CdTe QDs. In the presence of different metal ions, CdTe QDs show a appearance of new vibrational peak at ~1714 cm$^{-1}$ with a red shift of existing $v_{OH}$ stretching mode. In the presence of $X^+$ ions, some $X^+$ ions replace the H atom of hydroxyl group from the QD surface and make form $X^{n+}$:⁻OOC-CdTe complex structure. The appearance of the new peak at ~1714 cm$^{-1}$ may be due to the $X^{n+}$:⁻OOC-CdTe complex structure formation in presence of $X^+$ ions which is further verified by the red shift and decrement in intensity of the $v_{OH}$ stretching mode at 2300 cm$^{-1}$ (Figure 8b-f).

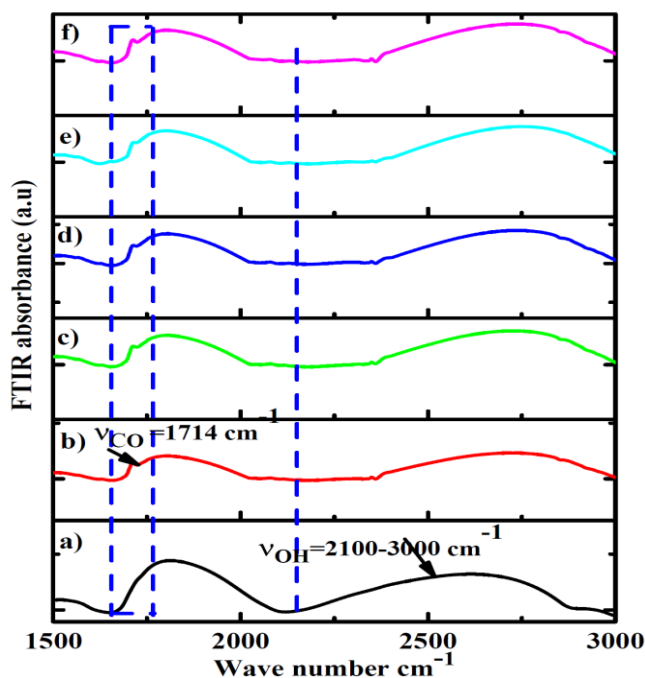

**Figure 8: a)** FTIR spectrum of COOH-CdTe QDs in hydroxylic medium and (**b-f**) FTIR spectra of $X^{n+}$: COOH-CdTe complexs in presence of $Cr^{3+}$, $Cu^{2+}$, $Pb^{2+}$, $Co^{2+}$, $Zn^{2+}$ metal ions in hydroxylic medium.

### *3.6 Analysis of real samples*

The practical applicability of COOH-CdTe QDs as metal ion sensor has been studied with different real life samples. To check the concentration of different dissolved metal ions, we have collected water samples from different areas of our society mainly urban river water (near to leather, chemical industry), rainwater in industrial area and normal paint water. Water samples were filtered through Whatman filter paper. Presence of different water dissolved metal ions in the real water samples were analysed by principal component analysis (PCA) and also by fluorescence analysis. PCA is an unsupervised statistical technique that maximizes the ratio of between-class and within-class variances and so allows the response patterns to be differentiated [67]. In PCA, the fluorescence intensity can be converted into a new set of linearly uncorrelated principal components as canonical factors like PC1, PC2 etc.For 2D analysis, two canonical factors (PC1, PC2) are observed to be more significant by the linear combination of the response matrices formed by observed fluorescence quantum yields of probe system. We have created 1st PCA analysis which again validated LOD value in Table1. CdTe QD in presence of known variable concentration of different metal ions ($Pb^{2+}$, $Cr^{3+}$, $Cu^{2+}$, $Zn^{2+}$ and $Co^{2+}$). The first PC (PC1, eigen value=6.46) accounts for 89.4% of the variability in the data. The second PC(PC2, eigen value=1.093) has 10.6% variance. So only the two PCs have the statistical significance according to Kaiser rule [68]. Since the first two PCs contributes 100% of the total variability so only two PCs (PC1, PC2) are used to plot the PCA scatter graph (Figure 9) for known concentration. The data in figure 9 (SD5) shows a distinct fluorescence response fingerprints for individual targets which helps us to detect and identify $Pb^{2+}$, $Cr^{3+}$, $Cu^{2+}$, $Zn^{2+}$ and $Co^{2+}$ions in solution phase accurately.

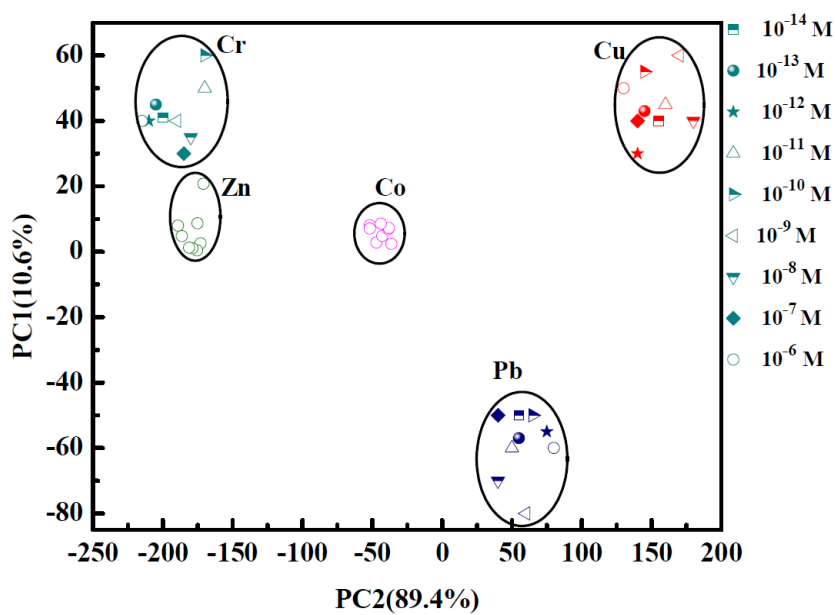

**Figure 9**: PCA score plot for $Cr^{3+}$, $Pb^{2+}$, $Cu^{2+}$, $Zn^{2+}$, and $Co^{2+}$ metal ion.

The PCA score plot represents a clear clustering of different concentrations of water dissolved metal ions ($Pb^{2+}$, $Cr^{3+}$, $Cu^{2+}$, $Zn^{2+}$ and $Co^{2+}$) for the concentration limit between $10^{-14}$ M to $10^{-6}$ M with no evident overlap between the present known metal ions (Figure 9). The variable concentrations of different metal ions present in solution can definitely affect the sensing ability of sensor. So this functionalized CdTe sensor is further investigated for its sensing ability with variable concentration limit from $10^{-14}$ M to $10^{-6}$ M. Nine sets of PC score data were generated from PCA processing for nine different known concentrations of specific metal ion and plotted them in one PCA scatter graph for that single metal ion. This clear clustering of concentration data determine the limit of detection (LOD) for our proposed senor as $10^{-14}$ M for $Pb^{2+}$, $Cr^{3+}$ and $Cu^{2+}$ metal ions. Evidence of clear and individual clustering for a specific metal ion clearly indicates the good sensitivity of functionalized

CdTe nanosensor to be used as a metal ions sensor. We have further tested the sensitivity of our proposed sensor with real samples by using the PCA scatter graph (Figure 10) by identifying the presence of specific metal ion and its tentative concentration. For this purpose we have compared the reference PCA scatter graph (Figure 9) which was created with known concentrations of metal ions with that of PCA scatter graph of the unknown real samples (SD7). For river water (ganga), the PCA score plot (Figure 10) (SD 6) confirms the presence of $Cr^{3+}$, $Cu^{2+}$, $Pb^{2+}$ $Co^{2+}$ and $Zn^{2+}$ ions in river water sample with a concentration of ~ $10^{-6}$ M , ~ $10^{-6}$ M, $1.28\times10^{-7}$M, ~$10^{-6}$ M and ~ $10^{-6}$M respectively.

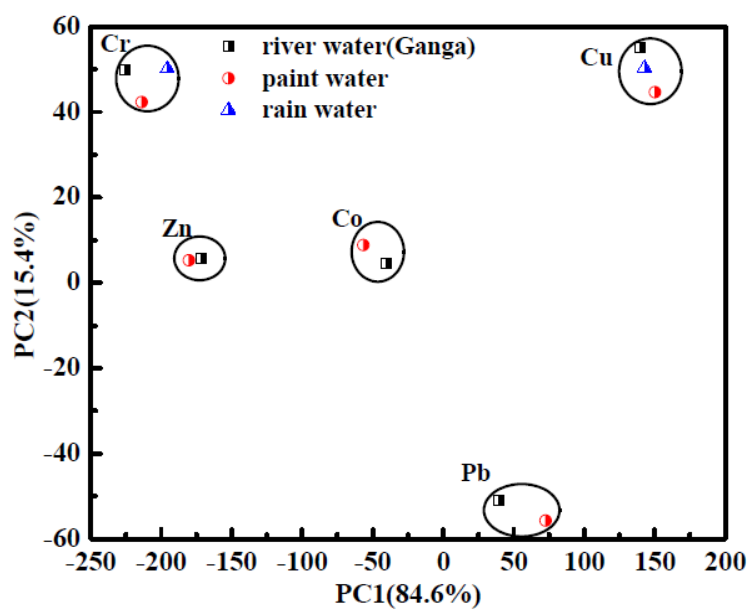

**Figure 10**: PCA score plot for $Cr^{3+}$, $Pb^{2+}$, $Cu^{2+}$, $Zn^{2+}$, and $Co^{2+}$ metal ion in Ganga river water, Paint, Rain water.

The observed concentrations of $Cr^{3+}$ and $Pb^{2+}$ ions are definitely beyond the permissible limit according to WHO guideline (Table 2) [22]. Similarly for industrially used normal paint water, the PCA scatter graph confirms the strong presence of $Pb^{2+}$ ions with its hazardous limit of $0.12\times10^{-6}$ M concentration (Table 2). For rain water in industrial area PCA scatter graph has confirmed the presence of only $Cr^{3+}$ and $Cu^{2+}$ ions with their very low concentration limit of ~ $10^{-11}$ M and ~ $10^{-12}$ M.

Sensitivity of our probe CdTenanosensorwas again verified by fluorescence analysis (SD 8). The decrement and red shift in the fluorescence peak intensity of functionalized CdTenanosensor in presence of different water dissolved metal ions present in real samples (river water, normal paint, rain water) have helped us to detect the present concentration of a specific metal ion in sample under observation. For river water, the presence of $Cr^{3+}$, $Cu^{2+}$, $Pb^{2+}$ $Co^{2+}$ and $Zn^{2+}$ ions have been confirmed with their concentration limit of $0.2\times10^{-6}$ M , $0.2\times10^{-9}$ M, $1.28\times10^{-7}$M, and $0.6\times10^{-6}$ M and $0.12\times10^{-6}$M respectively (Table 2). Similarly the probe sensor confirms the presence of $Cu^{2+}$ and $Pb^{2+}$ ions with concentration limit $1.12\times10^{-8}$ M and $0.2\times10^{-6}$ M for normal paint. Rain water also shows decrement in fluorescence intensity and shift in fluorescence peak position which confirms the presence of $Cr^{3+}$ and $Cu^{2+}$ with a concentration of $4.8\times10^{-11}$ M and $3.2\times10^{-12}$ M (Table 2). So from the evidence of both PCA analysis and emission data we got almost same sensitivity for our proposed CdTe nanosensor for different metal ions in real samples which perfectly matches with the existing WHO reported data [22]. Moreover, for the first time the proposed functionalized CdTe nanosensor has detected the existence of very low concentration($\sim 10^{-12}$) of $Cu^{2+}$ ion in real sample which is far better than other reported data [42].

|  | Real samples | $Cr^{3+}$[M] | $Cu^{2+}$[M] | $Pb^{2+}$[M] | $Zn^{2+}$[M] | $Co^{2+}$[M] |
|---|---|---|---|---|---|---|
| **PCA analysis** | Ganga river water | $\sim 10^{-6}$ | $\sim 10^{-6}$ | $1.28\times10^{-7}$ | $\sim 10^{-6}$ | $\sim 10^{-6}$ |
|  | Paint | $\sim 10^{-12}$ | $1.12\times10^{-8}$ | $0.12\times10^{-6}$ | $\sim 10^{-6}$ | $\sim 10^{-6}$ |
|  | Rain water | $\sim 10^{-11}$ | $\sim 10^{-12}$ | - | - | - |
| **Emission spectra analysis** | Ganga river water | $0.2\times10^{-6}$ | $0.2\times10^{-6}$ | $1.28\times10^{-7}$ | $0.6\times10^{-6}$ | $0.12\times10^{-6}$ |
|  | Paint | - | $1.12\times10^{-8}$ | $0.2\times10^{-6}$ | - | - |
|  | Rain water | $4.8\times10^{-11}$ | $3.2\times10^{-12}$ | - | - | - |
| **Permissible limit as per WHO data[22]** |  | $1.8\times10^{-7}$ | $1.9\times10^{-5}$ M | $4.8\times10^{-8}$ M | $7.6\times10^{-4}$ | - |
| **Metal ions in river water[22]** |  | $7.8\times10^{-6}$ | $2.2\times10^{-5}$ | $4.8\times10^{-7}$ | $7.6\times10^{-5}$ | - |

**Table 2**: Concentration for $Cr^{3+}$, $Pb^{2+}$, $Cu^{2+}$, $Zn^{2+}$, and $Co^{2+}$ metal ions present in river water, Paint, Rain water.

**4. Conclusion**

In the present study, we have investigated the applicability and selectivity of water soluble COOH-CdTeQDs as metal ion sensor. The synthesized QDs were characterized using UV–Vis absorption, emission and FTIR spectroscopic techniques. It is clearly observed that the surface coating by –COOH functional group increases the solubility and sensing activity of CdTe QDs. Variation of temperature and time influences the optical response of CdTe QDs. The present work shows the effective applicability of CdTe QDs as fluorescent nanosensor for different environmentally hazardous heavy metals ($Cr^{3+}$, $Pb^{2+}$, $Cu^{2+}$, $Zn^{2+}$, and $Co^{2+}$) in aqueous phase. Sensing efficiency of functionalized CdTe QDs is observed by its fluorescence "Turn-Off" response and also from PCA analysis for various water solvable metal ions having different concentrations ranging from $10^{-6}$M to $10^{-14}$M. Job plot analysis validates the existence of strong 1:1 complexation ($X^{n+}$: $^-$OOC-CdTe) between guest (COOH-CdTe QDs) and host ($X^{n+}$). The fluorescence intensity of CdTe QDs and the concentration of different metal ions ($X^{n+}$) have been described by Stern-Volmer equation. Strong appearance of $F-F_0/F_0$ establishes the better sensing capability of COOH-CdTe QDs for $Cr^{3+}$, $Pb^{2+}$, $Cu^{2+}$, $zn^{2+}$, and $Co^{2+}$ ions with their sensitive "Turn-Off fluorescent responses in aqueous environment which is far better than the existing fluorescent sensors available in industry. Due to its very low value of LOD, our probe functionalized CdTe QD nanosensor can detect efficiently a very low concentration (up to $10^{-14}$ M) of different water solvable metal ions. Exact match between PCA analysis and emission data have further verified the sensitivity and selectivity of proposed functionalized CdTe nanosensor. Again the proposed optical techniques of sensing were successfully applied for different real samples like: paint water, river water, rain water etc which again validates the applicability of COOH functionalized CdTe QD as a very effective opto chemical sensor for various water dissolved metal ions with variable concentrations. With its very efficient and low detection limit value for different water dissolved hazardous metal ions, our proposed COOH-CdTe nanosensor has definitely established its strong candidature to be used as raw material for the formation of a good sensory device.